\documentclass[12pt]{article}
\usepackage{epsf}
\title{ {\bf $\Delta\pi N$ coupling constant in light cone QCD \\
sum rules}}
\author{ \vspace{1cm} \\ 
 A. Gokalp\thanks{agokalp@metu.edu.tr}~ and~
 O. Yilmaz\thanks{oyilmaz@metu.edu.tr} 
\\
Physics Department, Middle East Technical University, \\
06531 Ankara, Turkey }
\date{\today}
\begin{document}
\setlength{\baselineskip}{24pt}
\maketitle
\setlength{\baselineskip}{7mm}
\begin{abstract}
We employ the light cone QCD sum rules to calculate $\Delta\pi N$
coupling constant by studying the two point correlation function between
the vacuum and the pion state. 
Our result is consistent with the traditional QCD sum rules calculations
and it is in agreement with the experimental value.
\end{abstract}
\thispagestyle{empty}
\newpage
\setcounter{page}{1}
\section{Introduction}
The determination of hadronic parameters, in particular meson-baryon coupling
constants, requires some information about the physics at large distances.
Therefore one has to employ some specific nonperturbative method to obtain
detailed predictions. Among the various nonperturbative methods,
QCD sum rules \cite{R1} have proved to be very useful to extract the low-lying
hadron masses and coupling constants. This method is a framework which connects
hadronic parameters with QCD parameters. It is based on short-distance
operator product expansion (OPE) in the deep Euclidean region of vacuum-vacuum
correlation function in terms of quark and gluon condensates. Further
progress has been achieved by an alternative method \cite{R2} known as the
QCD sum rules on the light cone.
The light cone QCD sum rules method is based on OPE on the light cone
which is an expansion over the twist of the operators rather than
the dimensions of the operators as in the traditional QCD sum rules approaches. In this expansion,
the main contribution comes from the lowest twist operators. The matrix elements of
non-local composite operators between meson and vacuum states define the
meson wave functions. The light-cone QCD sum rules have been employed to study
hadronic properties and these application can be found in \cite{R3}-\cite{R10}
and references therein.

In this work we study the $p\pi^{+}\rightarrow\Delta^{++}$ transition and
we employ the light-cone QCD sum rules approach to calculate the
$\Delta\pi N$ coupling $g_{\Delta\pi N}$. 
We note that this coupling constant was investigated by using the traditional 
QCD sum rules method in the soft pion limit \cite{R11}.
In the spirit of the philosophy
of the light cone QCD sum rules we study the time-ordered two-point
correlation function of the interpolating field for $N$ and $\Delta$
with pion. We use a dispersion relation that relates the hadronic
spectral density to the correlation function and thus write the
correlator in terms of phenomenological hadronic spectrum. Since the
interpolating fields are written in terms of quark fields, we also
calculate this correlator directly from QCD on the light cone. We consider
the structure $q_{\mu}$ and match these two descriptions. In order
to calculate the coupling constant  $g_{\Delta\pi N}$, we invoke
the double Boorel transformation so that the excited states and the
continuum contributions can be safely seperated out.

We consider the two-point correlation function with pion
\begin{eqnarray}
\Pi_{\mu}(p,q)=\int d^{4}x~e^{ipx}~<0|T\{\eta^{\Delta}_{\mu}(x)\bar{\eta}^{N}(0)\}|\pi^{+}(q)>
\label{E1}
\end{eqnarray}
with $p$ and ${\eta}^{\Delta}_{\mu}$ the four-momentum and the interpolating
current of delta, ${\eta}^{N}$ the interpolating current of nucleon, and $q$
the four-momentum of pion. The interpolating currents for delta
and nucleon \cite{R12} are
\newpage
\begin{eqnarray}
{\eta}^{\Delta}_{\mu}&=&\epsilon_{abc}(u_{a}^{T}C\gamma_{\mu}u_{b})u_{c}
\label{E2} \\
{\eta}^{N}&=&\epsilon_{abc}(u_{a}^{T}C\gamma_{\mu}u_{b})\gamma_{5}\gamma^{\mu}d_{c} \label{E3}
\end{eqnarray}
where a,b,c are the color indices, $C=i\gamma_{2}\gamma_{0}$ is the charge conjugation matrix,
T denotes transpose, and $u$ and $d$ are up and down quark fields respectively.

In order to construct the phenomenological side of the correlator we use
the following Lagrangian density
\begin{eqnarray}
{\cal{L}}=g_{\Delta\pi N}\bar{N}\Delta_{\mu}\partial^{\mu}\pi \label{E4}
\end{eqnarray}
which defines the $\Delta\pi N$ coupling constant $g_{\Delta\pi N}$ and
where $N$, $\Delta_{\mu}$ and $\pi$ are the nucleon, delta and pion fields.
Thus, at the phenomenological level the correlator can be saturated
by delta and nucleon states as
\begin{eqnarray}
\Pi_{\mu}(p,q)=
\frac{<0|\eta^{\Delta}_{\mu}|\Delta><\Delta|N\pi><N|\bar{\eta}^{N}|0>}
                    {(p^{2}-m_{\Delta}^{2})(p^{\prime}~^{2}-m_{N}^{2})}+...
                    \label{E5}
\end{eqnarray}
with the contributions from the higher states and the continuum denoted by dots. The overlapping
amplitudes of the interpolating currents with delta and nucleon states are
\begin{eqnarray}
<N|\bar{\eta}^{N}|0>=\lambda_{N}\bar{u}^{N}(p^{\prime}) \label{E6} \\
<0|\eta^{\Delta}_{\mu}|\Delta>=\lambda_{\Delta}u^{\Delta}_{\mu}(p) \label{E7}
\end{eqnarray}
where $u^{\Delta}_{\mu}$ is the Rarita-Schwinger spinor
and the matrix element $<\Delta|N\pi>$ is given as
\begin{eqnarray}
<\Delta|N\pi>=-g_{\Delta\pi N}~q^{\nu}~\bar{u}^{\Delta}_{\mu}(p)u^{N}(p^{\prime})~~~~~~~.
\label{E07} \end{eqnarray}
Substitution of  Eqs.(6,7,8) into Eq. (5) results in 
\begin{eqnarray}
\Pi_{\mu}(p,q)=-\frac{\lambda_{N}\lambda_{\Delta}g_{\Delta\pi N}}
           {(p^{2}-m_{\Delta}^{2})(p^{\prime}~^{2}-m_{N}^{2})}
&[&g_{\mu\nu}-\frac{1}{3}\gamma_{\mu}\gamma_{\nu}-\frac{2p_{\mu}p_{\nu}}{3m_{\Delta}^{2}}
+\frac{\gamma_{\mu}p_{\nu}-\gamma_{\nu}p_{\mu}}{3m_{\Delta}}]
\nonumber \\
&&~~~~~\times(\rlap/p+m_{\Delta})q^{\nu}(\rlap/p^{\prime}+m_{N})+... \label{E9}
\end{eqnarray}
We then consider the theoretical part of the correlator Eq. (1), for which
we obtain
\begin{eqnarray}
\Pi_{\mu}(p,q)=\int d^{4}x~e^{ipx}~[-2(S(\gamma_{\nu}C)S^{T}(C\gamma_{\mu})
{\cal{M}}(\gamma^{\nu}\gamma_{5})
-{\cal{M}}(\gamma^{\nu}\gamma_{5})tr((C\gamma_{\mu})S(\gamma_{\nu}C)S^{T})]
\label{E10}
\end{eqnarray}
with
\begin{eqnarray}
{\cal{M}}&=&\gamma_{5}<0|\bar{d}_{a}(0)\gamma_{5}u_{a}(x)|\pi>
-\gamma_{5}\gamma_{\lambda}<0|\bar{d}_{a}(0)\gamma_{5}\gamma^{\lambda}u_{a}(x)|\pi>
\nonumber \\
&&+\frac{1}{2}\sigma_{\alpha\beta}<0|\bar{d}_{a}(0)\sigma^{\alpha\beta}u_{a}(x)|\pi>~.
\label{Eq11}
\end{eqnarray}
The full light quark propagator in Eq. (10) with both perturbative term and
contribution from vacuum fields is given as
\begin{eqnarray}
iS(x)=&&i\frac{\rlap/x}{2\pi^{2}x^{4}}-\frac{<\bar{q} q>}{12}
-\frac{x^{2}}{192}<\bar{q} g_{s}~\sigma\cdot G~ q> \nonumber \\
&&-i\frac{g_{s}}{16\pi^2}\int_{0}^{1}du~\{\frac{\rlap/x}{x^2}\sigma\cdot
G(ux)
-4iu\frac{x_{\mu}}{x^{2}}G^{\mu\nu}(ux)\gamma_{\nu}\}+... \label{E12}
\end{eqnarray}
The matrix elements of the nonlocal operators between the vacuum and pion
state defines the two particle pion wave functions, and up to twist four the
Dirac components of these wave functions can be written as \cite{R13}
\begin{eqnarray}
<0|\bar{d}(0)\gamma_{\mu}\gamma_{5}u(x)|\pi^+>&=&if_{\pi}q_{\mu}\int_{0}^{1}du~e^{-iqux}
(\varphi_{\pi}(u)+x^{2}g_{1}(u)) \nonumber \\
&&+f_{\pi}(x_{\mu}-\frac{x^{2}q_{\mu}}{q^{2}}) \int_{0}^{1}du~e^{-iqux}g_{2}(u)~, \label{E13} \\
<0|\bar{d}(0)i\gamma_{5}u(x)|\pi^+>&=&\frac{f_{\pi}m_{\pi}^{2}}{m_{u}+m_{d}}
\int_{0}^{1}du~e^{-iqux}\varphi_{P}(u)~~~, \label{E14} \\
<0|\bar{d}(0)\sigma^{\mu\nu}\gamma_{5}u(x)|\pi^+>&=&(q_{\mu}x_{\nu}-q_{\nu}x_{\mu})
\frac{if_{\pi}m_{\pi}^{2}}{6(m_{u}+m_{d})}
\int_{0}^{1}du~e^{-iqux}\varphi_{\sigma}(u)~. \label{E15}
\end{eqnarray}
We further define
\begin{eqnarray}
G_{2}(u)=-\int_{0}^{u}g_{2}(v)dv, ~~~~G_2(1)=G_2(0)=0 \label{E16}
\end{eqnarray}
and
\begin{eqnarray}
g_{3}(u)=g_{1}(u)+G_{2}(u)~~~~~~~~~~. \label{E17}
\end{eqnarray}
From the theqretical part of the correlator given in Eq. (10),
we consider the structure $q_{\mu}$, and for this structure after
Fourier transformation over $x$ and double Borel transformations with respect to
variables $p_{1}^{2}=p^{2}$ and $p_{2}^{2}=(p-q)^{2}$ we finally obtain
\begin{eqnarray}
\Pi^{theor.}&=&f_{\pi}\mu_{\pi}<\bar{q} q>[~2M^{2}f_{0}(s_{0}/M^{2})
                  -\frac{1}{2}m_{0}^{2}~]~\varphi_{P}(u_{0})u_{0} \nonumber \\
           &&+\frac{f_{\pi}}{\pi^{2}}~[-\frac{1}{2}M^{6}f_{2}(s_{0}/M^{2})
                 +\frac{1}{24}g_{s}^{2}<G^{2}>M^{2}f_{0}(s_{0}/M^{2})~]~
                                            \varphi_{\pi}(u_{0}) \nonumber \\
           &&+\frac{f_{\pi}}{\pi^{2}}~[-\frac{1}{6}M^{6}f_{2}(s_{0}/M^{2})
                 -\frac{1}{24}g_{s}^{2}<G^{2}>M^{2}f_{0}(s_{0}/M^{2})~]~
                                u_{0}\varphi_{\pi}^{\prime}(u_{0}) \nonumber \\
           &&+\frac{f_{\pi}}{\pi^{2}}~\frac{1}{6}~g_{s}^{2}<G^{2}>g_{2}(u_{0})~
                  \frac{f_{\pi}}{\pi^{2}}~f_{1}(s_{0}/M^{2})g_{3}(u_{0})
\nonumber \\
           &&+\frac{f_{\pi}}{\pi^{2}}~[~2M^{4}f_{1}(s_{0}/M^{2})
                 +\frac{1}{6}g_{s}^{2}<G^{2}>]~u_{0}~g_{3}^{\prime}(u_{0})
\nonumber \\
           &&+\frac{1}{18}f_{\pi}\mu_{\pi}<\bar{q} q>
                  [-8M^{2}f_{0}(s_{0}/M^{2})\varphi_{\sigma}(u_{0})
+(-4M^{2}f_{0}(s_{0}/M^{2})+m_{0}^{2})u_{0}\varphi_{\sigma}^{\prime}(u_{0})]
\nonumber \\
&& \label{E18}
\end{eqnarray}
where the function
\begin{eqnarray}
f_{n}(s_{0}/M^{2})=1-e^{-s_{0}/M^{2}}
{\sum_{k=0}^{n}}\frac{(s_{0}/M^{2})^{k}}{k!} \nonumber
\end{eqnarray}
is the factor used to subtract the continuum, which is modeled by
the dispersion integral in the region $s_{1},~s_{2}\geq s_{0}$, $s_{0}$
being the continuum threshold,
$\mu_{\pi}=\frac{m_{\pi}^{2}}{m_u+m_d}$, and
\begin{eqnarray}
u_{0}=\frac{M_{1}^{2}}{M_{1}^{2}+M_{2}^{2}},~~~~
M^{2}=\frac{M_{1}^{2}M_{2}^{2}}{M_{1}^{2}+M_{2}^{2}} \nonumber
\end{eqnarray}
with $M_{1}^{2}$ and $M_{2}^{2}$ are the Borel parameters, and
$\varphi^{\prime}(u_{0})=\frac{d\varphi}{du}\mid_{u_{0}}$.

Performing double Borel transformation over the variables $p_{1}^{2}=p^{2}$
and $p_{2}^{2}=(p-q)^{2}$ on the phenomenological part in Eq. (9), and then
equating the result obtained for the Lorentz structure $g_{\mu\nu}$ part
to that theoretical result given in Eq. (18) we finally obtain the sum rule
for the coupling constant $g_{\Delta\pi N}$
\begin{eqnarray}
g_{\Delta\pi N}\lambda_{N}\lambda_{\Delta}=
\frac{2}{(m_{N}+m_{\Delta})^{2}}e^{\frac{m_{\Delta}^{2}}{M_{1}^{2}}}
e^{\frac{m_{N}^{2}}{M_{2}^{2}}}\Pi^{theor.}~~. \label{E19}
\end{eqnarray}
We note that this sum rule depends on the value of the wave functions
at a specific point, which are much better known than the whole wave
functions \cite{R14}.

\section{Numerical results and discussion}

The various parameters we adopt are
$m_{0}^{2}=0.8~ GeV^{2}$,
$<g_{s}^{2}G^{2}>=0.474~GeV^{4}$,
$f_{\pi}=0.132~GeV$,
$\mu_{\pi}=1.65~GeV$,
$<\bar{q}q>=-(0.225~GeV)^{3}$,
$s_{0}=3~GeV^{2}$.
In our calculation of the theoretical part we use the two particle
pion wave functions based on the QCD sum rule approach given in
\cite{R14} as
\begin{eqnarray}
\varphi_{\pi}(u,\mu)&=&
6u\bar{u}~[~1+a_{2}(\mu)C_{2}^{3/2}(2u-1)+a_{4}C_{4}^{3/2}(2u-1)~]~, \nonumber \\
\varphi_{\sigma}(u,\mu)&=&
6u\bar{u}~\big [~1+C_{2}\frac{3}{2}~[~5(u-\bar{u})^{2}-1~]
+C_{4}\frac{15}{8}~[~21(u-\bar{u})^{4}-14(u-\bar{u})^{2}+1~]~\big ]~, \nonumber \\
\varphi_{P}(u,\mu)&=&
1+B_{2}\frac{1}{2}~[~3(u-\bar{u})^{2}-1~]
+B_{4}\frac{1}{8}~[~35(u-\bar{u})^{4}-30(u-\bar{u})^{2}+3~]~, \nonumber \\
g_{1}(u,\mu)&=&
\frac{5}{2}\delta^{2}(\mu)u^{2}\bar{u}^{2}+\frac{1}{2}\epsilon (\mu)
~[~u\bar{u}(2+13u\bar{u})+10u^{3}\ln u~(2-3u+\frac{6}{5}u^{2})
\nonumber \\
&&+10\bar{u}^{3}\ln \bar{u}~(2-3\bar{u}+\frac{6}{5}\bar{u}^{2})~]~, \nonumber \\
g_{2}(u,\mu)&=&\frac{10}{3}\delta^{2}u\bar{u}(u\bar{u})~, \nonumber \\
G_{2}(u,\mu)&=&\frac{5}{3}\delta^{2}u^{2}\bar{u}^{2}~,
\label{E20}
\end{eqnarray}
where $\bar{u}=1-u$, $C_{2}^{3/2}$ and $C_{4}^{3/2}$ are the Gegenbauer
polynomials defined as
\begin{eqnarray}
C_{2}^{3/2}(2u-1)&=&\frac{3}{2}~[~5(2u-1)^{2}+1~]~, \nonumber \\
C_{4}^{3/2}(2u-1)&=&\frac{15}{8}~[~21(2u-1)^{4}-14(2u-1)^{2}+1~]~,
\label{E21}
\end{eqnarray}
and $a_{2}=2/3$, $a_{4}=0.43$ corresponding to the normalization point
$\mu=0.5~GeV$. The remaining parameters are taken from the QCD sum rule
estimates of Ref. \cite{R15} as $\delta^{2}(\mu=1~GeV)=0.2$, and
$\epsilon (\mu=1~GeV)=0.5$ and from those of Refs.\cite{R3,R13} as
$B_{2}=0.48$, $B_{4}=1.51$, $C_{2}=0.10$, $C_{4}=0.07$.
Since the mass difference of $N$ and $\Delta$ is not very significant,
we let $M_{1}^{2}=M_{2}^{2}=2M^{2}$ from which it follows that $u_{0}=1/2$.

We study the dependence of the sum rule of Eq. (19) on the continuum threshold
$s_{0}$ and on the Borel parameter $M^{2}$. We find that the sum rule is
very stable with resonable variations of  $s_{0}$ and $M^{2}$ as can be seen
in Fig. 1. In order to determine the value of the coupling constant
$g_{\Delta\pi N}$ from the sum rule Eq. (19), the residues
$\lambda_{\Delta}$ and $\lambda_{N}$ of the hadronic currents are needed.
We use the following values that are obtained from the corresponding sum rules for
$\Delta$ and $N$ which are \cite{R11}
\begin{eqnarray}
|\lambda_{N}|^{2}~2(2\pi)^{4}~e^{-\frac{m_{N}^{2}}{M^{2}}}&=&
M^{6}f_{2}(s_{0}^{N}/M^{2})+b~M^{2}f_{0}(s_{0}^{N}/M^{2})+\frac{4}{3}a^{2}~,
\label{E22} \\
|\lambda_{\Delta}|^{2}~5(2\pi)^{4}~e^{-\frac{m_{\Delta}^{2}}{M^{2}}}&=&
M^{6}f_{2}(s_{0}^{\Delta}/M^{2})
-\frac{25}{18}~b~M^{2}f_{0}(s_{0}^{\Delta}/M^{2})+\frac{20}{3}a^{2}~,
\label{E23}
\end{eqnarray}
where
\begin{eqnarray}
a&=&-2\pi^{2}<\bar{q}q>=0.5~GeV^{3}~, \nonumber \\
b&=&\frac{\alpha_{s}}{\pi}<G^{2}>=0.12~GeV^{4}~, ~~and \nonumber \\
s_{0}^{\Delta}&=&(m_{\Delta}+0.5)^{2}~,~~~s_{0}^{N}=(m_{N}+0.5)^{2}~.
\nonumber
\end{eqnarray}
We substitute these values of the residues into the sum rule in Eq. (19),
and further study the dependence of $g_{\Delta\pi N}$ on the continuum
threshold $s_{0}$ and on the Borel parameter $M^{2}$. This dependence
is shown in Fig. 2, examination of which indicates that it is quite
stable with resonable variations of $s_{0}$ and $M^{2}$.
We thus obtain the coupling constant $g_{\Delta\pi N}$ as
\begin{eqnarray}
g_{\Delta\pi N}=(14.5\pm 1.5)~ GeV^{-1} \label{E24}
\end{eqnarray}
The uncertainty we included is due to the variation of the continuum
threshold and the Borel parameter.

Finally, we consider the Lagrangian density in Eq. (4) and obtain the
expression for the decay width as
\begin{eqnarray}
\Gamma (\Delta^{++}\rightarrow \pi^{+}p)=\frac{1}{6}\frac{g_{\Delta\pi N}^{2}}{4\pi}
\frac{(m_{N}+m_{\Delta})^{2}-m_{\pi}^{2}}{m_{\Delta}^{2}}~p_{\pi}^{3}
\label{E25}
\end{eqnarray}
where the pion momentum $p_{\pi}$ is
\begin{eqnarray}
p_{\pi}=\sqrt{(\frac{m_{\pi}^{2}+m_{\Delta}^{2}-m_{p}^{2}}{2m_{\Delta}})^{2}-m_{\pi}^{2}}~~.
\nonumber
\end{eqnarray}
We use the experimental result \cite{R16} for the decay width and obtain the coupling constant
as
\begin{eqnarray}
g_{\Delta\pi N} =(15.2\pm 0.1)~~~GeV^{-1}   \nonumber
\end{eqnarray}
which indicates that our value in Eq. (24) is in satisfactory agreement with the experimental 
result.

In summary, we calculated $\Delta\pi N$ coupling constant
$g_{\Delta\pi N}$ using light-cone QCD sum rules. Our result is consistent with
the results obtained using traditional QCD sum rules \cite{R11} and it 
is in good agreement with the value of the coupling constant deduced from the
experimental decay rate of  $\Delta^{++}$ baryon.

\vspace{0.5cm}
{\bf Acknowledgment}\\
We thank to T. M. Aliev for suggesting us this problem and
his guidance during the course of our work.

\pagebreak

\newpage

{\bf Figure Captions:}

\begin{description}

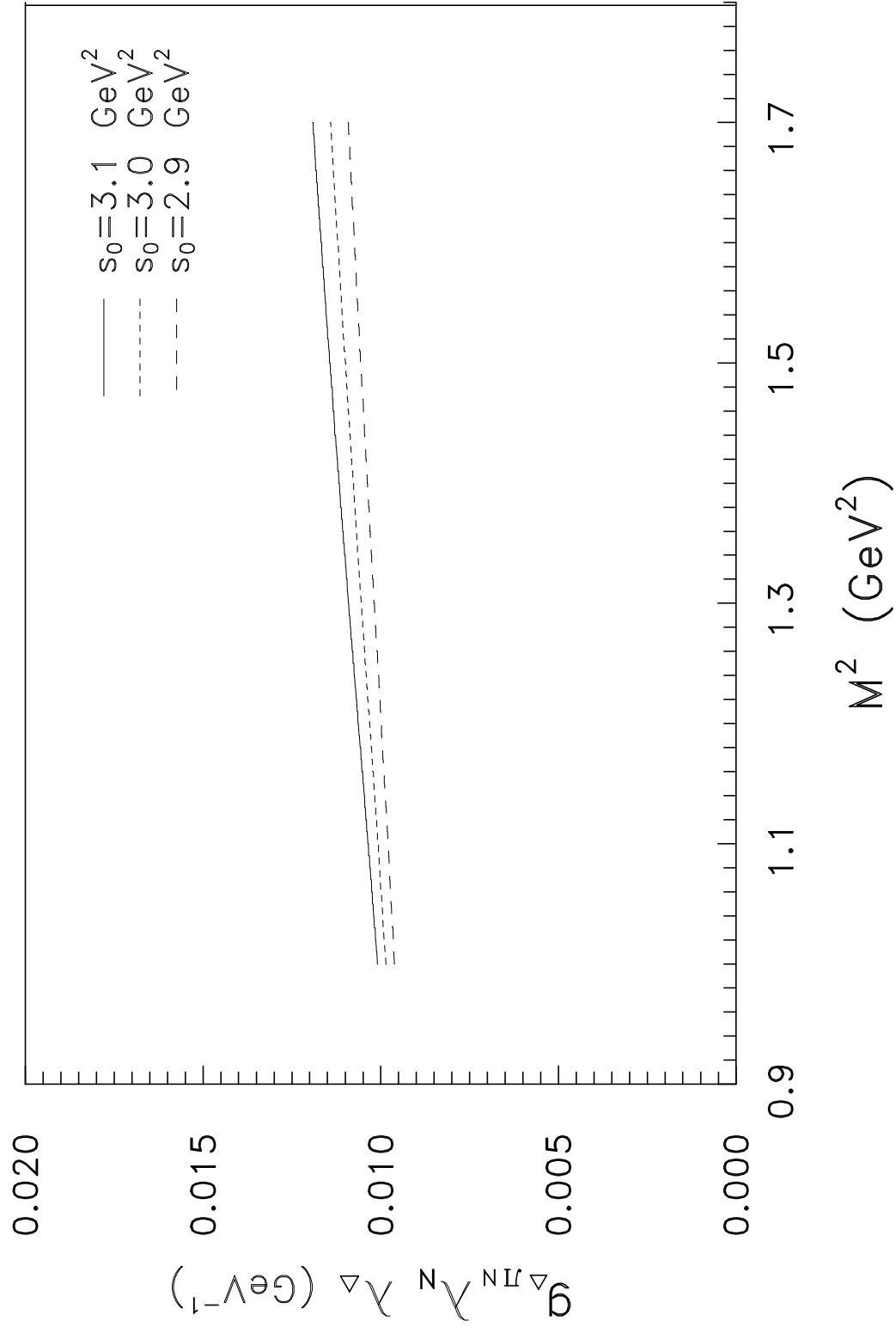
\item[{\bf Figure 1}:] The dependence of $g_{\Delta\pi N}\lambda_{\Delta}\Lambda_{N}$
on the Borel parameter $M^{2}$ and on the continuum threshold $s_{0}$.

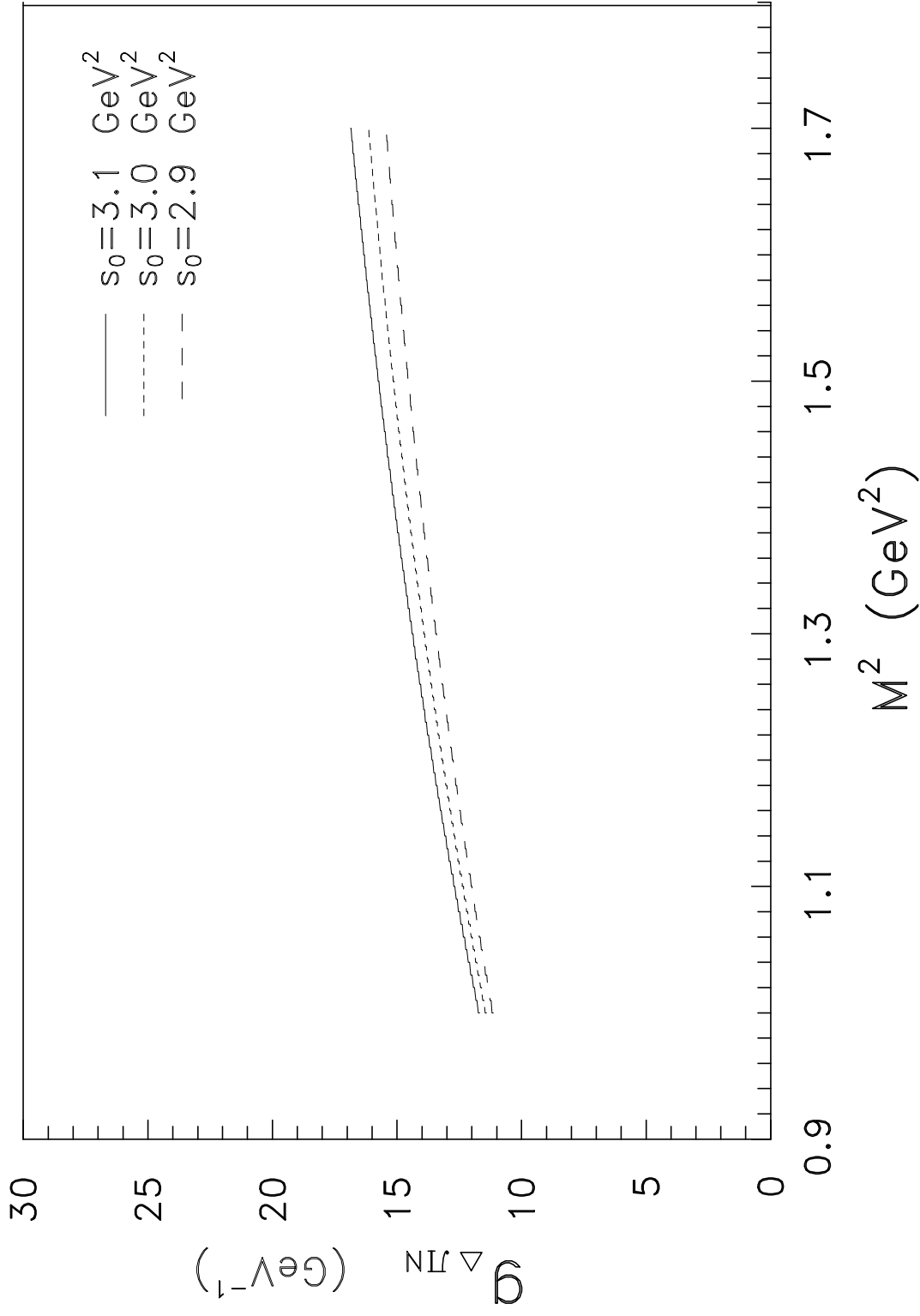
\item[{\bf Figure 2}:] The sum rule for $g_{\Delta\pi N}$ as a function of the
 Borel parameter $M^{2}$ and the continuum threshold $s_{0}$.

\end{description}

\end{document}